\documentclass[12pt]{revtex4}
\usepackage{dcolumn}
\usepackage{bm}
\usepackage{epsfig}

\def\be{\begin{equation}}
\def\bea{\begin{eqnarray}}
\def\ee{\end{equation}}
\def\eea{\end{eqnarray}}

\newcommand{\up}{\uparrow}

\newcommand{\dn}{\downarrow}
\newcommand{\ek}{\epsilon_\bfk}
\newcommand{\cks}{c_{\bfk,\sigma}}

\newcommand{\cku}{c_{\bfk,\uparrow}}

\newcommand{\cmkd}{c_{-\bfk,\downarrow}}

\newcommand{\bz}{b_0}

\newcommand{\jm}{{ J}_-}
\newcommand{\jp}{{ J}_+}
\newcommand{\jx}{{ J}_x}
\newcommand{\jy}{{ J}_y}
\newcommand{\jz}{{ J}_z}

\newcommand{\bfk}{{\bf k}}

\def\pt{\partial}

\def\dt{\delta}

\def\eps{\varepsilon}

\def\const{\mbox{const}}

\def\mod{\mbox{ mod}}

\def\ln{\mbox{ln}}

\begin{document}

\title{Change in the adiabatic invariant in a nonlinear two-mode model of
Feshbach resonance passage. }

\vskip 10mm


\vskip 5mm

\author{A.P. Itin$^{1,2}$,  A.A. Vasiliev$^2$, G.Krishna$^1$, and S.
Watanabe$^1$.} \affiliation{$^1$Department of Applied
Physics and Chemistry, University of Electro-Communications,\\
1-5-1, Chofu-ga-oka, Chofu-shi, Tokyo 182-8585, Japan\\ $^2$Space
Research Institute, Russian Academy of
Sciences,\\
Profsoyuznaya str. 84/32, 117997 Moscow,
Russia}\email{alx_it@yahoo.com}


\begin{abstract}
Mean-field approach has recently been used to model coupled
atom-molecular Bose-Einstein condensates (BEC) and coupled
Fermi-Bose condensates near Feshbach resonance. Sweeping of
magnetic field across the resonance gives a new (nonlinear)
version of Landau-Zener problem. We investigate the structure of
the corresponding classical phase space and calculate change in
the action which corresponds to finite-rate efficiency of the
sweep. We consider the case of non-zero initial action, which
corresponds to some finite initial molecular fraction.
\end{abstract}

\maketitle

 \pagebreak

\section{Introduction}

Adiabatic invariance is very important issue in quantum mechanics
\cite{Ehrenfest,Navarro}. Relation between slow quantum
transitions and change in the adiabatic invariant of a linear
oscillator has been studied in \cite{Dykhne}. Dynamics of
Bose-Einstein condensates \cite{E1,E2,E3,E4,GP,Pit} introduces a
paradigm of nonlinearity into quantum systems. In many mean-field
models related to BEC physics (like  nonlinear Landau-Zener models
\cite{Zobay,IW0}, macroscopic quantum self-trapping \cite{Smerzi},
etc), nonlinear effects that are common to classical nonlinear
systems have been revealed. A conceptual phenomenon of classical
adiabatic theory is destruction of the adiabatic invariance at
separatrix crossings \cite{AKN} which is encountered in different
fields of physics. It is of great importance for BEC physics
because change in the classical action of a nonlinear two-state
model corresponds to probability of transition between the two
states (modes). Here we consider a nonlinear mean-field model of a
slow sweep through a Feshbach resonance in a quantum gas of
fermionic atoms coupled to BEC of diatomic molecules \cite{TPBFV1}
(for brevity, we call this system a Bose-Fermi condensate). A
number of closely related non-stationary problems have come up
recently in context of coupled Bose-Fermi condensates
(\cite{Altshuller,Barankov,TPBFV1}) and coupled atom-molecular BEC
\cite{Links,Milburn,Yurovsky}). The mean-field approach to such
problems is very interesting and not at all trivial
\cite{Altshuller}. In \cite{Altshuller} the following Hamiltonian
describing a fermion-boson condensate is considered:

\be \hat H=\sum_{j,\sigma}\eps_{j} \hat c_{j \sigma }^\dagger \hat
c_{j \sigma}+\omega \hat b^\dagger \hat b +g\sum_j \left( \hat
b^\dagger \hat c_{ j\dn} \hat c_{ j\up}+ \hat b \hat
c_{j\up}^\dagger \hat c_{j\dn}^\dagger\right),  \label{fermibose}
\ee

where $\eps_j$ are the single-particle energy levels and the
operators $\hat c_{j \sigma }^\dagger$ ($\hat c_{j \sigma}$)
create  (annihilate)  a fermion of one of the two species
$\sigma=\up$ or $\dn$  in an orbital eigenstate of energy
$\eps_j$. In case the  single-particle potential  is
translationally invariant, $|j\up\rangle=|{\bf k}\up\rangle$ and
$|j\dn\rangle=|-{\bf k}\dn\rangle$ \cite{Altshuller}. Operators
$\hat b^\dagger$ ($\hat b$) create (annihilate) quanta of the
bosonic field.

The mean-field approximation of  (\ref{fermibose}) amounts to
treating the bosonic field classically, i.e. replacing operators
$\hat b^\dagger$ and $\hat b$ with c-numbers in the Heisenberg
equations of motion. This procedure is justified provided the
bosonic mode is macroscopically populated. As shown in
\cite{Altshuller}, in this approximation the dynamics of
(\ref{fermibose}) coincides with that of a classical Hamiltonian
system (where classical dynamical variables are the time-dependent
quantum-mechanical expectation values $\langle \hat c_{ j\dn} \hat
c_{ j\up}\rangle$, $\langle \hat b\rangle$, and
$\langle\sum_\sigma \hat c_{j \sigma }^\dagger \hat c_{j
\sigma}\rangle$ ). The mean-field approximation is also equivalent
to replacing operators with classical variables and their
commutators with Poisson brackets. It is also shown in
\cite{Altshuller} that the fermion-oscillator model
(\ref{fermibose}) can be viewed as a generalization of the Dicke
(Tavis-Cummings \cite{TC}) model of Quantum Optics:  the latter
model corresponds to zero fermionic bandwidth limit of
(\ref{fermibose}), i.e. to the case when all single-particle
levels $\eps_j$ are degenerate, $\eps_j= \mu$. To demonstrate
this, \cite{Altshuller} reformulates (\ref{fermibose}) as a
spin-oscillator model using Anderson's pseudospins (i.e., $2\hat
K_j^z=\hat n_j-1, \qquad \hat K_j^-=\hat c_{j\dn} \hat
c_{j\up},\qquad \hat K_j^+=\hat c_{j\up}^\dagger \hat
c_{j\dn}^\dagger $, where $\hat n_j=\sum_\sigma \hat c_{j \sigma
}^\dagger \hat c_{j \sigma}$), and  obtains the Hamiltonian

\be \hat H=\sum_{j=0}^{n-1} 2\eps_j \hat K_j^z+\omega \hat
b^\dagger \hat b+ g\sum_{j=0}^{n-1} \left( \hat b^\dagger\hat
K_j^- + \hat b\hat K_j^+\right). \label{dicke} \ee

In the zero bandwidth ("degenerate") limit the Hamiltonian
(\ref{dicke}) reduces to the Dicke model

\be \hat H_{Dicke}=2\mu\hat T_z+\omega \hat b^\dagger \hat b+
g(\hat b^\dagger \hat T_-+\hat b \hat T_+), \label{TC} \ee
describing an interaction of a single collective spin $\hat {\bf
T}=\sum_j \hat {\bf K}_j$ with a harmonic oscillator. Mean-field
solution of (\ref{TC}) was discussed in \cite{dicke,BP}. The more
general many-body problem (\ref{fermibose},\ref{dicke}) also turns
out to be integrable. Explicit solutions for the mean-field
dynamics of the model (\ref{fermibose},\ref{dicke}) were
constructed in \cite{Barankov}. Later, comprehensive solutions for
the mean-field dynamics were obtained in \cite{Altshuller} using a
method of separation of variables
\cite{komarov,sklyanin,kuznetsov}, which allowed to derive a
complete set of integrals of motion for
(\ref{fermibose},\ref{dicke}). Quantum solutions of
(\ref{fermibose}) can be obtained by the Bethe ansatz
\cite{Zhou,Duk}. In \cite{Links2}, quantum model (\ref{fermibose})
and  a more general version which includes s-wave scattering
interactions were solved using the boundary quantum inverse
scattering method (QISM) as developed by Sklyanin
\cite{sklyanin2}; interesting enough, through the exact solution,
the spectrum can be mapped into that of a Schrodinger equation.

In the present paper, we deal only with the mean-field dynamics.
The mean-field solutions of \cite{Altshuller} describe dynamics of
the system that has been prepared in a nonequilibrium state at
$t=0$. To model Feshbach resonance passage converting Fermi atoms
to Bose molecules, one may use the model (\ref{fermibose}) with
time-dependent $\omega$. For example, in \cite{TPBFV1}, the model
(\ref{fermibose}) with zero fermionic bandwidth and time-dependent
molecular energy $ $ was considered using another approach
(similar to that of \cite{Yurovsky}). To be more precise,
\cite{TPBFV1} considers the Hamiltonian $ H=\sum_{\bfk,\sigma}
\ek\cks^\dag\cks+{\cal E}(t) \bz^\dag\bz +g\left(\sum_{\bf k}
\cku\cmkd \bz^\dag+H.c.\right)~,$ with $\ek=\hbar^2k^2/2m$ being
the kinetic energy of an atom with mass $m$, in the degenerate
limit where $\ek = {\bf \eps}$ for all $\bfk$. Introducing
operators $ \jm=\frac{\bz^\dag\sum_{\bfk} \cku\cmkd}{(N/2)^{3/2}}
~,~ \jp=\frac{\sum_{\bfk} \cmkd^\dag\cku^\dag \bz}{(N/2)^{3/2}}, $
$ \jz=\frac{\sum_{\bfk,\sigma}\cks^\dag\cks-2\bz^\dag\bz}{N}, $
where $N=2\bz^\dag\bz+\sum_{\bfk,\sigma}\cks^\dag\cks$ is the
conserved total number of particles,  one gets (after certain
rescalings) \cite{TPBFV1} the Heisenberg equations of motion for
the association of a quantum-degenerate gas of fermions,
\begin{eqnarray}
\frac{d}{d\tau}\jx &=& -\delta(\tau)\jy \nonumber \\
\frac{d}{d\tau}\jy &=& \delta(\tau)\jx+\frac{3\sqrt{2}}{4}
\left(\jz-1\right)\left(\jz+\frac{1}{3}\right)
-{\sqrt{2}\over{N}}\left(1+\jz\right), \nonumber \\
\label{HVardi} \frac{d}{d\tau}\jz&=&\sqrt{2}\jy,
\end{eqnarray}
which depend on the single parameter $\delta(\tau) = ({\cal
E}(t)-2 {\bf \eps} )
 /\sqrt{N}g$. Mean-field limit of (\ref{HVardi}) is obtained by
 replacing $J_{x,y,z}$ with their expectation values $u,v,w$, and
 omitting the quantum-noise term ${\sqrt{2}\over{N}}\left[1+\jz\right]$ (which is justified since the mean-field approximation
 is valid only up to terms of order $1/\sqrt{N}$; the mean-field approach to Feschbach resonance passage is discussed in \cite{Ja,Ju}). The resulting system
 of equations ($\dot u = -\dt(\tau)v,  \quad
\dot v = \dt(\tau)u + \frac{\sqrt 2}{4} (w-1)(3w+1), \quad \dot w
= \sqrt 2 v $) is analyzed in detail in the present paper (similar
models arise in two-mode approximation for coupled atom-molecular
BEC \cite{Links,Milburn,Yurovsky};  also, a three-mode model
considered in \cite{Links3} at certain conditions has very similar
phase portraits).

We consider the model of \cite{TPBFV1}, concentrating on the case
of non-zero initial molecular fraction. Within the model, change
in the action at the resonance passage gives the remnant atomic
fraction as a power-law of a sweeping rate parameter (instead of
the exponential law \cite{Landau} of Landau-Zener linear model ).
Classical adiabatic theory \cite{AKN} provides a method to
calculate change in the actions in nonlinear systems. It is a
well-known fact that action is an adiabatic invariant in a
Hamiltonian system that depends on a slowly varying parameter.
This result is based on the possibility of averaging over fast
motion in the unperturbed [frozen at a certain parameter value]
system. The situation is somehow different if the unperturbed
system has separatrices on its phase portrait. As the parameter
varies, the separatrices slowly evolve on the phase portrait. In
particular, the area surrounded by a separatrix may change. Hence,
a phase trajectory of the exact system may cross the separatrix of
the frozen system. On the separatrix, the period of motion is
equal to infinity. This results in breakdown of the averaging
method and in this case more accurate study is necessary to
describe behavior of the action. It turns out that at the crossing
a quasi-random jump in the value of the adiabatic invariant
occurs. The asymptotic formula for this jump in a Hamiltonian
system depending on a slowly varying parameter was obtained in
\cite{T,CET,N86}. Later, the theory of adiabatic separatrix
crossings was also developed for slow-fast Hamiltonian systems
\cite{N87}, volume preserving systems \cite{NV99}, and was applied
to certain physical problems (see, for example,
\cite{VVN1,INV00,ILNV02}). In this paper we use the methods of
\cite{N86} to obtain a variation of the adiabatic invariant at the
separatrix crossing. A separate paper \cite{IW0} considers four
different nonlinear two-mode models related to BEC physics,
concentrating on classical non-adiabatic phenomena, and predicts
new nonlinear effects there. Section II of the present paper
introduces the model and discusses the structure of the classical
phase space, while Section III contains calculations of the change
of the adiabatic invariant at the resonance passage.

\section{Main equations and phase portraits}

We consider the classical Hamiltonian

\be H = \dt(\tau) w + (1-w)\sqrt{1+w} \cos \phi \label{2.4} \ee
with a slowly varying parameter $\dt$. Such model appears as a
mean-field limit of two-mode approximation of coupled atomic and
diatomic-molecular Bose-Einstein condensates
\cite{Links,Milburn,Yurovsky} and in the mean-field treatments of
coupled degenerate gas of Fermi atoms and BEC of molecules, as
discussed in the Introduction. In the latter case, Feshbach
resonance passage can be modelled by sweeping the parameter
$\delta$, and corresponding change in the classical action is
related to the remnant atomic fraction after the sweep.

\begin{figure*}
\includegraphics[width=80mm]
{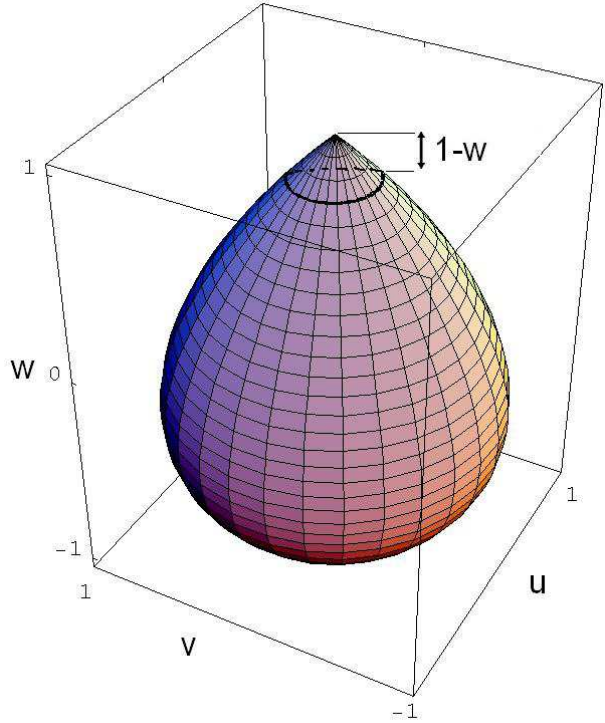}
 \caption{Geometry of the generalized Bloch sphere (the surface
 $u^2+v^2=\frac{1}{2}(w-1)^2(w+1)$).}
  \label{sphere}
\end{figure*}

The system can also be investigated using the equations of motion
for a generalized Bloch vector \cite{TPBFV1}:

\bea
\dot u &=& -\dt(\tau)v, \nonumber \\
\dot v &=& \dt(\tau)u + \frac{\sqrt 2}{4} (w-1)(3w+1),
\label{2.1}\\
\dot w &=& \sqrt 2 v, \nonumber \eea where the dot denotes time
derivative, $\tau = \eps t$, $0<\eps \ll 1$ is a small parameter,
and $\dt(\tau)$ is a slowly varying parameter corresponding to the
[scaled] detuning. Equations of motion (\ref{2.1}) restricted to
surface (\ref{2.2}) are equivalent to equations of motion in
Hamiltonian system with the Hamiltonian (\ref{2.4}), where
canonically conjugated variables are $w$ and $\phi$, $\phi =
\arctan(v/u)$.
\begin{figure*}
\includegraphics[width=120mm]
{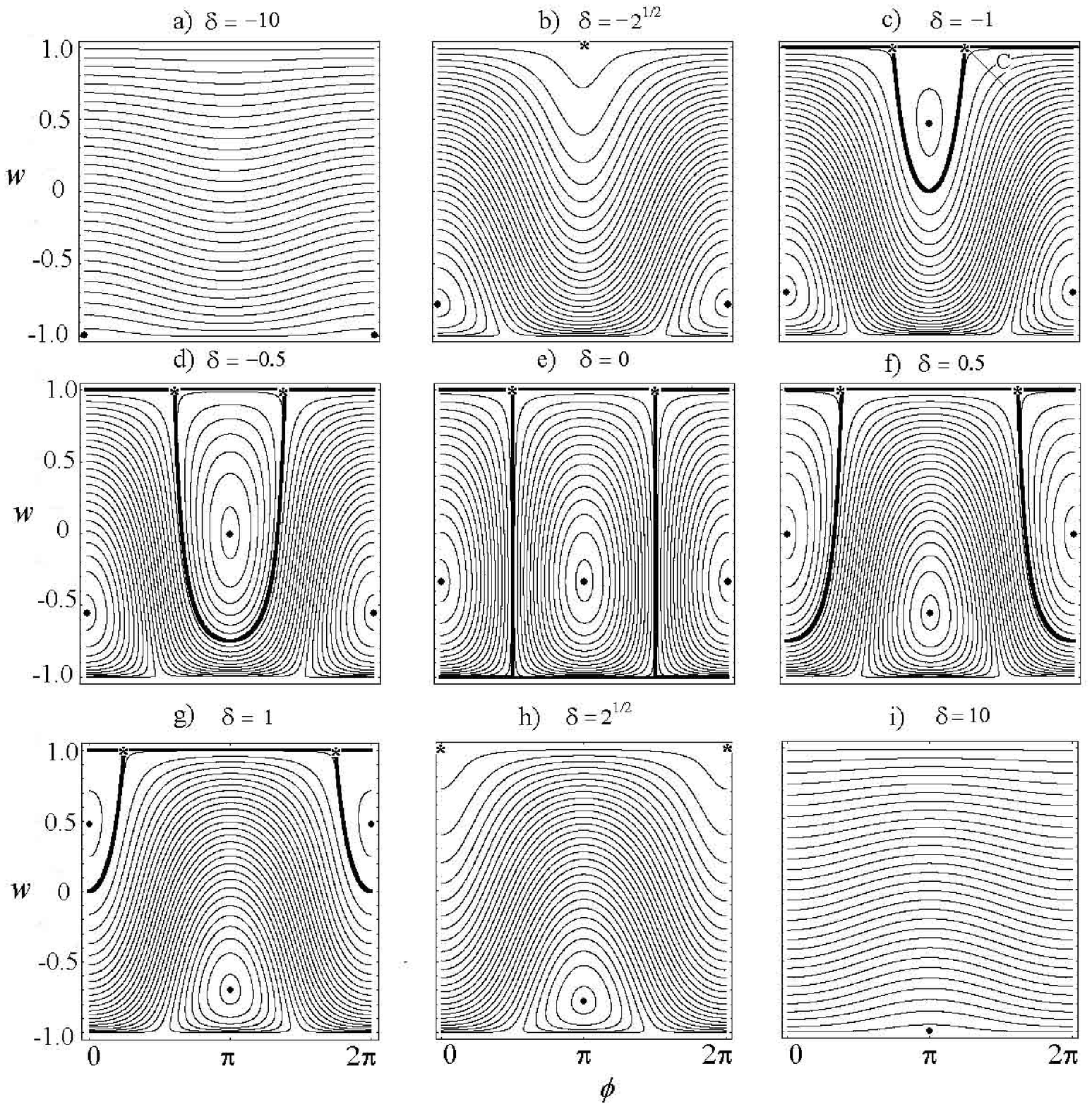}
 \caption{
Phase portraits of the system with Hamiltonian (\ref{2.4}) frozen
at different values of parameter $\delta$.}

\end{figure*}

First, we consider equations (\ref{2.1}) at frozen values of
parameter $\dt$. In this case, these equations possess two first
integrals. One of them is $u^2 + v^2 - \frac12 (w-1)^2(w+1) =
\const$. Corresponding to the conservation of single-pair
atom-molecule coherence, the constant should be taken equal to
zero. The equation

\be u^2 + v^2 - \frac12 (w-1)^2(w+1) = 0. \label{2.2} \ee defines
a surface of rotation around $w-$axis with a singular [conical]
point at (0,0,1) (see Fig.1). Another integral of (\ref{2.1}) at a
frozen value of $\dt$ is given by

\be u + \frac{\dt}{\sqrt 2} w = \const, \label{2.3} \ee and at
different values of the constant defines a family of planes
parallel to $v-$axis. The angle between these planes and $w-$axis
depends on the value of $\delta$. Hence, trajectories of frozen
system (\ref{2.1}) are given by intersections of surface
(\ref{2.2}) and planes (\ref{2.3}). At $|\dt|<\sqrt 2$, plane $u =
\dt (1-w)/\sqrt 2$, passing through the singular point (0,0,1)
defines a singular trajectory on surface (\ref{2.2}).

Consider phase portraits of the system with Hamiltonian
(\ref{2.4}) frozen at different values of parameter $\delta$ [see
Figure 2]. In these portraits, points (0,0,1) and (0,0,-1) of the
$(u,v,w)$ space are represented as segments $w=1$ and $w=-1$
correspondingly. Point (0,0,1) is always a stable point of
(\ref{2.1}), but formally speaking this is not true for the points
of the segment $w=1$ on phase portraits of the system with
Hamiltonian (\ref{2.4}).  If $\dt$ is negative and $|\dt|>\sqrt
2$, there is only one stable elliptic point on the phase portrait,
at $\phi = 0$ and $w$ not far from $-1$ [see Figure 2a]. It may
seem that there are also two singular points at $w=-1, \phi =
\pi/2, \; 3\pi/2$, but they are just an artefact of the
representation in variables $w,\phi$; the phase trajectory passing
through these points corresponds to the trajectory of (\ref{2.1})
passing through (0,0,-1), and the period of motion on this
trajectory is finite.

At $\dt = -\sqrt 2$ a bifurcation takes place, and at $0>\dt>
-\sqrt 2$ the phase portrait looks as shown in Figs.2c,d. There
are two saddle points at $w = 1, \cos\phi=\dt/\sqrt 2$ and a
newborn elliptic point at $\phi = \pi$. The trajectory connecting
these two saddles corresponds to the singular trajectory on the
surface (\ref{2.2}). On the phase portrait of the system with
Hamiltonian (\ref{2.4}) it separates rotations and oscillating
motions and we call it the separatrix of the frozen system. Below,
we will consider "shifted" Hamiltonian $E=H-\dt$; $E$ equals zero
on the separatrix.

At $\dt = 0$ the singular trajectory on the surface (\ref{2.2}) is
passing through (0,0,-1), and correspondingly on the phase
portrait the segment $w=-1$ belongs to the separatrix [Figure 2
e]. At $0<\dt< \sqrt 2$ the phase portrait looks as shown in Figs.
2 f,g. Finally, at large positive values of $\dt$, again there is
only one elliptic stationary point at $\phi = \pi$, and $w$ close
to $-1$.

It may be easier to understand the transition between Figure 2 h
and Figure 2 i using also the $(u,v,w)$ representation (see Figure
1). The elliptic point in Figure 2 h corresponds to the elliptic
point on the surface (\ref{2.2}). This latter elliptic point is
close to the pole $(0,0,-1)$ of the surface (\ref{2.2}). As
parameter $\dt$ grows, this elliptic point approaches to the pole.
Consider a closed trajectory of small enough diameter on the
surface (\ref{2.2}), surrounding this elliptic point. This
trajectory is given by an intersection of surface (\ref{2.2}) and
one of the planes (\ref{2.3}) defined by a certain value of the
constant. While $\dt$ is not large enough, this trajectory does
not embrace the pole. The corresponding phase trajectory in Figure
2 h surrounds the elliptic point. At $\dt$ large enough, the
trajectory on surface (\ref{2.2}) corresponding to the same value
of the constant in (\ref{2.3}) embraces the pole. Along the
corresponding phase trajectory in Figure 2 i, value of $\phi$
varies from $0$ to $2\pi$.

Consider now a phase trajectory on a phase portrait frozen at a
certain value of $\dt$. If the trajectory is closed, the area $S$
inside of it is connected with the action $I$ of the system by a
simple relation $S = 2\pi I$. If the trajectory is not closed, we
define the action as follows. If the area $S$ bounded by the
trajectory and lines $w=1, \phi=0, \phi=2\pi$ is smaller than
$2\pi$, we still have $S = 2\pi I$. If $S$ is larger than $2\pi$,
we put $2\pi I = 4\pi - S$. Defined in this way, $I$ is a
continuous function of the coordinates.

In system with Hamiltonian (\ref{2.4}) with $\tau = \eps t$, the
action $I$ is an adiabatic invariant of motion (see, for example,
\cite{AKN}). Far from the separatrix, it undergoes oscillations of
order $\eps $.

In the following, we  will use the so called improved adiabatic invariant $J= I +\eps
\tilde u(w,\phi,\tau)$ rather than $I$. The function $u$ is defined as follows [see, for
example, \cite{N86}]:

\be
\tilde u = \tilde u(w,\phi,\tau) = \frac{1}{2\pi} \int ^T _0 \left( \frac{T}{2} - t
\right) \frac{\partial E}{\partial \tau} d t.
\label{2.5}
\ee
The integral in (\ref{2.5}) is computed along the
unperturbed trajectory passing through the point $(w,\phi)$ at the
time moment $t=0$; $T$ is the period of motion on this trajectory.
If the trajectory is not closed, $T$ is the time necessary for a
phase point on this trajectory to cover the distance of $2\pi$
along the $\phi$-axis. [Definition (\ref{2.5}) is valid provided
that on this trajectory the action $I$ and the area on the phase
portrait are related simply as $S=2\pi I$; in the following
calculations of the jump in the adiabatic invariant, we need to
consider only such trajectories.] Far from the separatrices,
variation of $J$ along a phase trajectory is of order $\eps ^2$.

As the slow time $\tau$ grows, the area bounded by the separatrix
$S^*(\tau)$ slowly changes. On the other hand, a value of the
adiabatic invariant associated with a certain phase trajectory
stays well-preserved. Accordingly, phase trajectories of
(\ref{2.4}) can cross the separatrix. Let initially $\dt$ be large
in magnitude and negative, so that the phase portrait is similar
to one shown in Figure 2a. Consider a trajectory rotating close to
the singular point on surface (\ref{2.2}). Along the corresponding
phase trajectory on the plane $(\phi,w)$, value of $w$ is close to
1. We assume $1-w$ on this trajectory to be small, yet finite.
Hence, the initial value $J_-$ of improved action $J$ is also
small. As the time goes, value of parameter $\dt$ grows, and at
$\dt=-\sqrt 2$ the separatrix loop appears. The area $S^*(\tau)$
surrounded by the separatrix grows with time, and the action
associated with the phase trajectory stays approximately constant.
In the so called improved adiabatic approximation this action is
conserved at $J=J_-$, and at the slow time moment $\tau=\tau_*$
such that $S^*(\tau_*)=2\pi J_-$ the phase trajectory crosses the
separatrix. Phenomena that take place at such crossing are
considered in the following section.

\section{Variation of the adiabatic invariant at the separatrix crossing}

First, consider the motion in the frozen system along a phase
trajectory close to the separatrix. Let on this trajectory $E=h,
\, |h|\ll 1$. [We consider such values of $\delta$ that $h$ is
positive outside the separatrix and is negative inside.] The main
part of the time a phase point on this trajectory spends in a
neighborhood of the saddle points at $w=1$. Linearizing the system
near the saddle points, we find that in the main approximation the
period of motion along this trajectory is

\be
T = - \frac{2}{\sqrt{2-\dt^2}} \ln |h| +b.
\label{3.1}
\ee
Here $b$ is a smooth function of $\dt$; its form is not important
for the rest of the argument. In agreement with the formula $T=
2\pi
\partial I/\partial h$ we find for the adiabatic invariant $I$ in the main
approximation

\be
2\pi I= S^* - \frac{2}{\sqrt{2-\dt^2}} h \ln |h| +(b +
\frac{2}{\sqrt{2-\dt^2}})h . \label{3.2}
\ee

Now compute the function $\tilde u$ at a point of the vertex bisecting the angle between
incoming and outgoing separatrices of the saddle point (see Figure 2c). From (\ref{2.5}),
(\ref{3.1}) one obtains in the main approximation:

\be
2\pi \tilde u = \frac{\Theta}{2\sqrt{2-\dt^2}} \ln |h| -b \frac{\Theta}{4} + d, \qquad
\Theta = \frac{\partial S^*}{\partial \tau}.
\label{3.3}
\ee
Here $d$ is a smooth function of $\dt$; its form is not important for the rest of the
argument. Consider now the separatrix crossing in the exact system with Hamiltonian $E$.
On the relevant interval of slow time we have $\Theta(\tau)>0$ [the area inside the
separatrix loop grows]. Initial values of the Hamiltonian and the improved adiabatic
invariant at $\tau = \tau_-$ are $h_-, J_-$. The phase point rotates close to the
separatrix. As the area bounded by the separatrix grows, the point comes closer to the
separatrix with each turn. Denote by $h_n, I_n, \tilde u_n$ values of the corresponding
functions at time moments $\tau _n$ when the trajectory crosses the vertex bisecting the
angle between incoming and outgoing separatrices of the saddle point $C$ outside the
separatrix loop (see Figure 2c). We enumerate $\tau _n$ as follows: $\tau_0$ is the time
of the last crossing of the vertex before crossing the separatrix, other $\tau_n$ have
negative numbers $\tau_0 > \tau _{-1}>...> \tau_{-N} \ge \tau_-$. Here $N \gg 1$ is a
large integer. Its exact value does not influence the result in the main approximation.
After crossing the separatrix, the phase point continues its motion inside the separatrix
loop. As the loop grows, the phase point goes deeper and deeper inside, rotating around
the elliptic point. At time moments $\tau_n, \, 1 \le n \le N$, the trajectory crosses
the vertex bisecting the angle between incoming and outgoing separatrices of the saddle
point $C$ inside the separatrix loop. The corresponding values of $h, I, \tilde u$ are
$h_n, I_n, \tilde u_n$.

It follows from (\ref{3.1}) - (\ref{3.3}), that in the main
approximation the following expressions are valid:

\bea
h_{n+1} &=& h_n - \eps \Theta, \nonumber
\\
\tau_{n+1} &=& \tau_n - \frac{\eps}{2\sqrt{2-\dt_*^2}} \left[ \ln
|h_n| + 3 \ln|h_n - \eps \Theta| \right]+ \eps b ,
\label{3.4}\\
2 \pi I_n &=& S^*(\tau_0) - \Theta (\tau_0 - \tau_n) - \frac{2}{\sqrt{2-\dt_*^2}} h_n \ln
|h_n| +(b + \frac{2}{\sqrt{2-\dt_*^2}})h , \nonumber
\\
2\pi \tilde u_n &=& \frac{\Theta}{2\sqrt{2-\dt_*^2}} \ln|h_n| -b \frac{\Theta}{4} + d.
\nonumber
\eea
Here and below the values of
$\Theta$, $b$, and $d$  are calculated at $\tau = \tau_*$, the
time of separatrix crossing in the adiabatic approximation,
$\dt_*$ is also the value of $\dt$ at $\tau = \tau_*$. Summing the
above expressions (\ref{3.4}) from $n=-N$ to $n=0$, we find the
change of the improved adiabatic invariant before the separatrix
crossing in the main approximation:

\be
2\pi (J_0 - J_{-N}) = \frac{\eps \Theta}{\sqrt{2-\dt_*^2}}
\left[ \ln \xi - 2 \xi \ln \xi + 2\xi -2 \ln (\sqrt{2\pi}/\Gamma
(\xi)) \right],
\label{3.5}
\ee
where $\xi = |h_0/(\eps \Theta)|$,
$\Gamma(\cdot)$ is the gamma function. At the separatrix crossing
we obtain:

\be
2\pi (J_1 - J_0) = \frac{\eps \Theta}{\sqrt{2-\dt_*^2}} \left[
-\ln \xi - \ln (1-\xi) + 2 (1-\xi) \ln (1-\xi) + 2\xi \ln\xi
\right].
\label{3.6}
\ee
For the change of $J$ after the
separatrix crossing we find

\be
2\pi (J_N - J_1) = \frac{\eps \Theta}{\sqrt{2-\dt_*^2}} \left[
\ln (1-\xi) - 2 (1-\xi) \ln (1-\xi) - 2\xi  -2 \ln
(\sqrt{2\pi}/\Gamma (1-\xi)) \right].
\label{3.7}
\ee

Far from the separatrix, the variation of $J$ is of order $\eps^2$
on time periods of order $1/\eps$. Hence, to obtain in the main
approximation the jump of $J$ at the separatrix crossing, one has
to sum up expression (\ref{3.5}) - (\ref{3.7}). Thus, we find:

\be
2\pi \Delta J = -2 \frac{\eps \Theta}{\sqrt{2-\dt_*^2}}
\ln(2\sin \pi \xi).
\label{3.80}
\ee
Note, that this result is similar to one obtained for a symmetrical double well (see
\cite{CET,N86}), though the geometry here is different. According to \cite{CET,N86}, the
error of formula (\ref{3.80}) is $O(\eps^{3/2} |\ln \eps|)$.

Formula (\ref{3.80}) can be simplified. The separatrix is defined by equation $E=0$.
Thus, the area inside the separatrix loop can be calculated as
\be
S^*(\delta) = 2 \int_{\delta^2 - 1}^1[ \pi- \arccos \left(
\frac{\delta}{\sqrt{1+w}}\right) ]\,d w
\ee
We are interested in derivative of $S^*(\delta)$ over $\delta$. Differentiating the above
integral over parameter $\delta$ one obtains:
\bea
\frac{\pt S^*}{\pt\delta} = - 2 \int_{\delta^2}^2 \frac{-dx}{\sqrt{x-\delta^2}}=
4\sqrt{2-\delta^2}
\eea
We have
\be
\Theta = \frac{\pt S^*}{\pt\delta} \cdot \delta^{\prime},
\ee
where $\delta^{\prime} \equiv \partial \delta/\partial \tau$. Therefore, formula
(\ref{3.80}) takes the following form:
\be
\Delta J = - \frac{4\eps \delta^{\prime}_*}{\pi}\, \ln( 2\sin \pi \xi).
\label{3.8}
\ee
Here $\delta^{\prime}_*$ is the value of $\delta^{\prime}$ calculated at time
$\tau=\tau_*$.

The value $\xi$ strongly depends on initial conditions: a small of order $\eps$ variation
of initial conditions results generally in variation of $\xi$ of order 1. Hence, this
value can be considered as a random variable; its distribution should be treated as
uniform on the segment $(0,1)$ [see \cite{N86}, \cite{N87}, see also \cite{INV00} for
numerically found distribution of $\xi$ in a similar problem]. Formula (\ref{3.8}) is
valid provided that $\xi$ is not too close to the ends of the interval $(0,1)$:
$k\sqrt\eps< \xi<1-k\sqrt\eps$, where $k$ is a positive constant, see \cite{N86}. The
value $\Delta J$ in (\ref{3.8}) should also be treated as random; we find its dispersion
below.

\begin{figure*}
\includegraphics[width=160mm]
{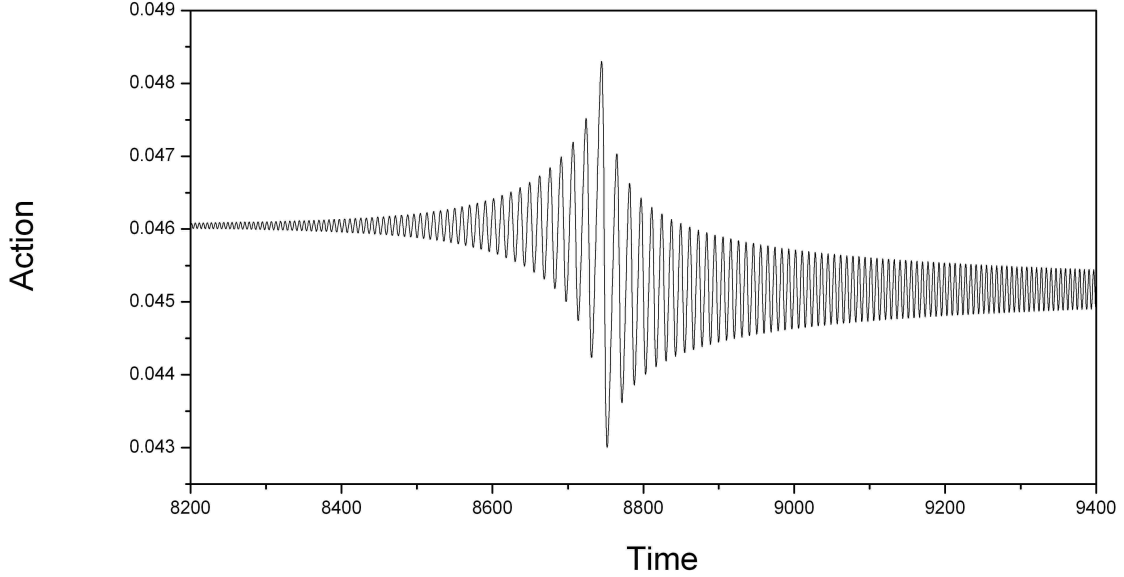}

 \caption{Typical jump of the adiabatic invariant (action) at separatrix
crossing.
  \label{jump}}
\end{figure*}

After the separatrix crossing, the phase point rotates around the elliptic point inside
the separatrix loop and slowly drifts with this point to the bottom of the phase
portrait. If $\dt(\tau)$ is a monotonous function, the phase trajectory will never again
cross the separatrix. Assume that $\dt(\tau)$ is a smooth function and $\dt = \dt_- =
\const$ at $\tau<\tau_-$, $\dt = \dt_+ = \const$ at $\tau>\tau_+$. [It is assumed that
$\dt_-<0,\, \dt_+>0$.] In other words, parameter $\dt$ is slowly monotonically varying
between two border values. Then $\pt E/\pt \tau = 0$ at $\tau = \tau_\pm$, and action $I$
coincides with $J$. Hence, formula (\ref{3.8}) gives the variation in action $I$. Let the
magnitudes of $\dt_-$ and $\dt_+$ be large enough. If the initial value of $w=w_-$ is
close enough to 1, the corresponding unperturbed phase trajectory is an almost straight
line (cf. Fig. 2 a). Hence, $I_-\approx (1-w_-)$. Similarly, $I_+\approx (-1-w_+)$. Thus,
variation in action $I$ corresponds to the remnant atomic fraction. In the adiabatic
limit, value of atom-molecular imbalance is reversed at the passage, while the change in
the action produce nonadiabatic correction to this result. A typical jump in the action
is shown in Fig.\ref{jump}.

\begin{figure*}
\includegraphics[width=100mm]
{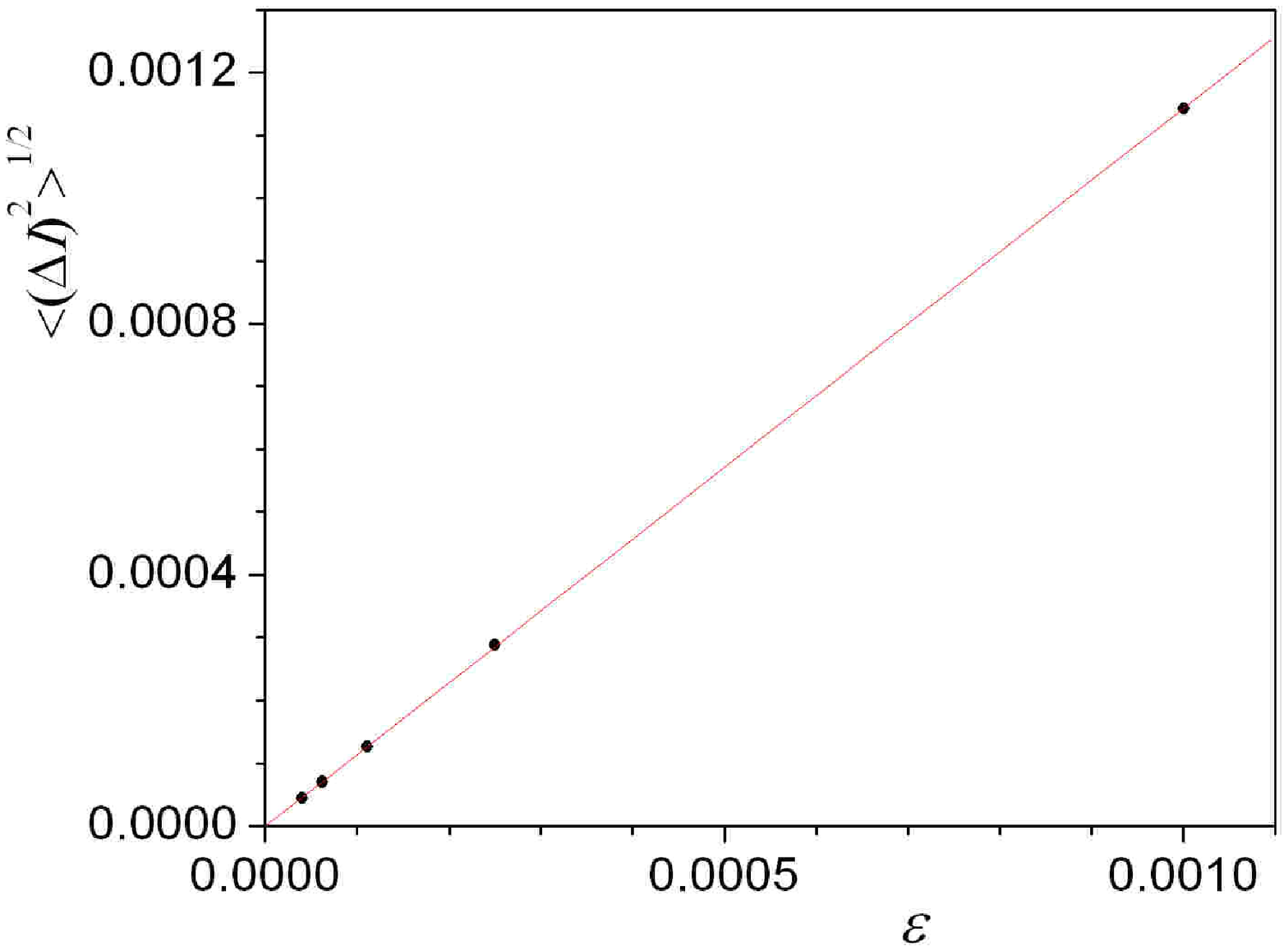}
 \caption{Scaling of jumps in improved adiabatic invariant with
 $\eps$. For each point on the plot, we take a set of 100 trajectories
with
 initial ( at $\delta=-10$) values of $w$ being closely
 distributed around $w=0.99=1-10^{-2}$. Final values of improved action were taken far from the
 separatrix (at $\delta=0$). The line on the plot is a linear fit to
data, and a perfect linear scaling can be seen. The slope of the line gives $\sigma
\approx 1.14 \eps$, and the coefficient is in good agreement with the theoretical value
$\sqrt{4/3} \approx 1.15$.
  \label{stat}}
\end{figure*}

Dispersion of jumps in the action can be predicted using formula (\ref{3.8}):

\be
\sigma^2 = 16 \eps^2 (\delta_*^{\prime})^2 \pi^{-2} \int_0^1 \ln^2 (2 \sin \pi \xi)\, d
\xi = \frac{4 \eps^2 (\delta_*^{\prime})^2}{3}
\ee
To check numerically the scaling of the jumps with $\eps$ and the dispersion, we
calculated bunches of trajectories with close initial conditions (see Fig.\ref{stat}).
For the numerical calculations, we used linear sweeping with $\delta^{\prime}=1$,
therefore the predicted value of dispersion is  $ \sigma^2 = (4/3) \eps^2$. Numerically
found coefficient is equal to $1.30$, which is in reasonable ( $2\%$ accuracy) agreement
with theoretical prediction $4/3= 1.3333$.

Consider now briefly the case when the external parameter $\delta$
varies slow periodically between a positive and a negative values
of large magnitudes. In this situation a phase trajectory crosses
the separatrix once on each period of variation of $\delta$. Each
crossing can be characterized by two values, namely $J$ and $\xi$.
Consider two subsequent crossings of the separatrix. Assume that
$J=J_1$ well before the first crossing, and the corresponding
value of $\xi$ is $\xi_1$. Let $J_2, \xi_2$ characterize the
second crossing. In the main approximation, we have the following
map:
\bea
J_2 &=& J_1 + \eps F(\xi_1)
\label{map1} \\
\xi_2 &=& \xi_1 + \eps^{-1} G(J_2)\; \mod \, 1,
\label{map2}
\eea
where
\bea F(\xi) = - \frac{\Theta}{\pi \sqrt{2-\dt_*^2}}
\ln(2\sin \pi \xi),
\nonumber \\
G(J) =\frac{1}{2\pi} \int_{\tau_1}^{\tau_2} \omega (J,\tau) \,
d\tau,
\nonumber
\eea
$\tau_1, \tau_2$ are values of the slow time
$\tau$ corresponding to the first and the second separatrix
crossings, calculated in the adiabatic approximation; $\omega$ is
a frequency of the fast motion. Suppose that the value $\xi_1$
gets a small variation $\Delta\xi_1$. According to (\ref{map1}),
this leads to variation of the jump in the improved adiabatic
invariant by a value $\eps F^{\prime}(\xi_1)\Delta\xi_1$, where
$F^{\prime}$ denotes derivative of function $F(\xi)$. Thus, as
follows from (\ref{map2}), the value $\xi_2$ obtains a variation
$\Delta\xi_2 = \Delta\xi_1 +
G^{\prime}(J_2)F^{\prime}(\xi_1)\Delta\xi_1$, and
\be
\frac{\Delta\xi_2}{\Delta\xi_1} - 1 \sim G^{\prime}F^{\prime}.
\label{criterium}
\ee
This latter value can be used to describe
the phase mixing in the system. If it is large, values $\xi_1$ and
$\xi_2$ are statistically independent. This is the case when
$\xi_1$ is close enough to $0$ or to $1$. Assume first that in the
process of iterations of the map (\ref{map1}), (\ref{map2}) all
the values $\xi_i$ are statistically independent. Then the
dynamics in the system results in diffusion of the adiabatic
invariant and mixing in the phase space. Indeed, the process can
be roughly modelled as a random walk in adiabatic invariant: the
time interval between two subsequent steps is of order
$\eps^{-1}$, and the length of each step is of order $\eps$.
Hence, total variation in $J$ after $N$ steps is $\sqrt{N}\eps$.
After $N\sim \eps^{-2}$ steps, which takes time of order
$\eps^{-3}$, the value of $J$ (and hence the remnant part of the
atomic fraction) will change by a value of order one.

\begin{figure*}
\includegraphics[width=0.4\linewidth]
{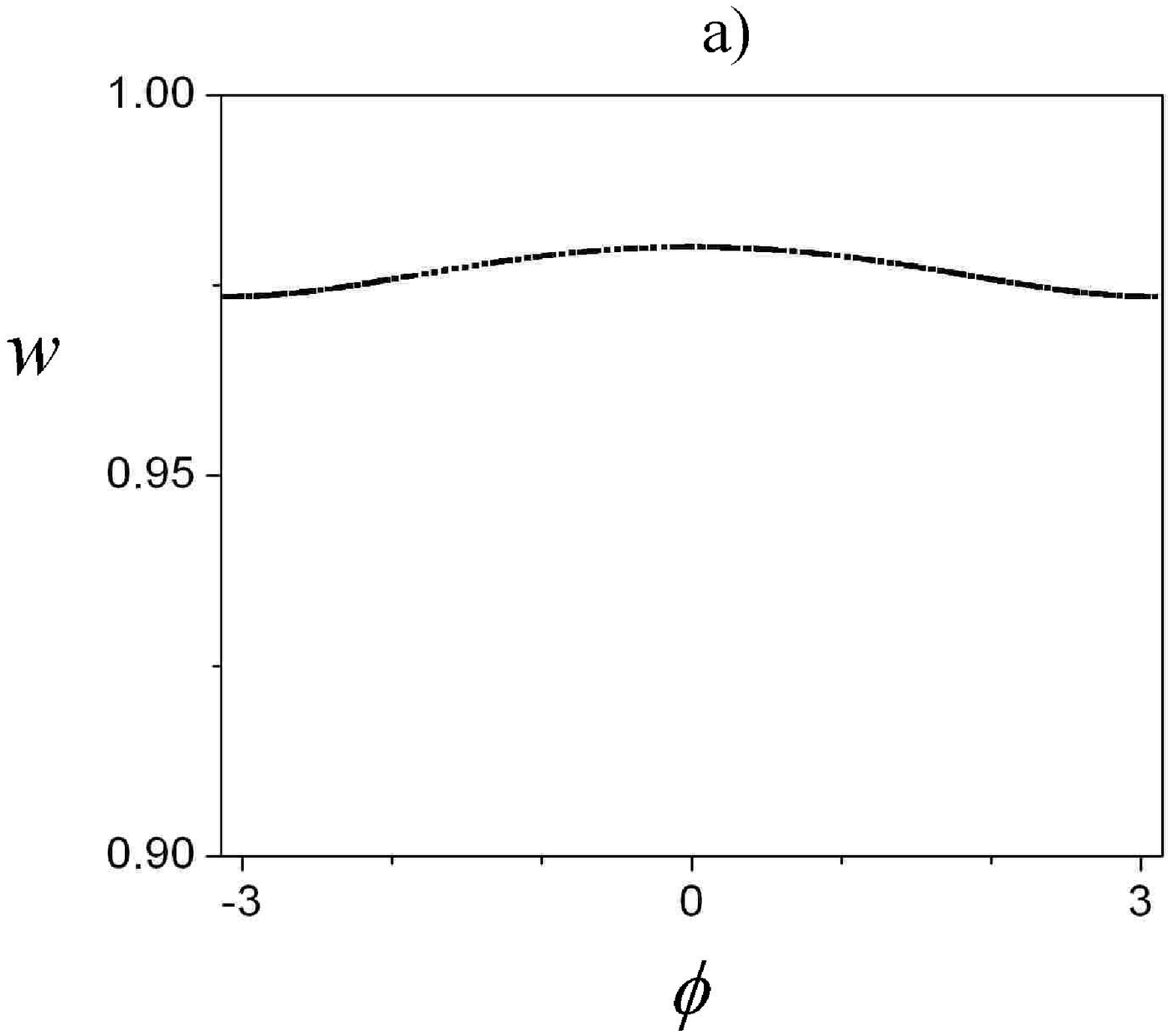}
\includegraphics[width=0.4\linewidth]
{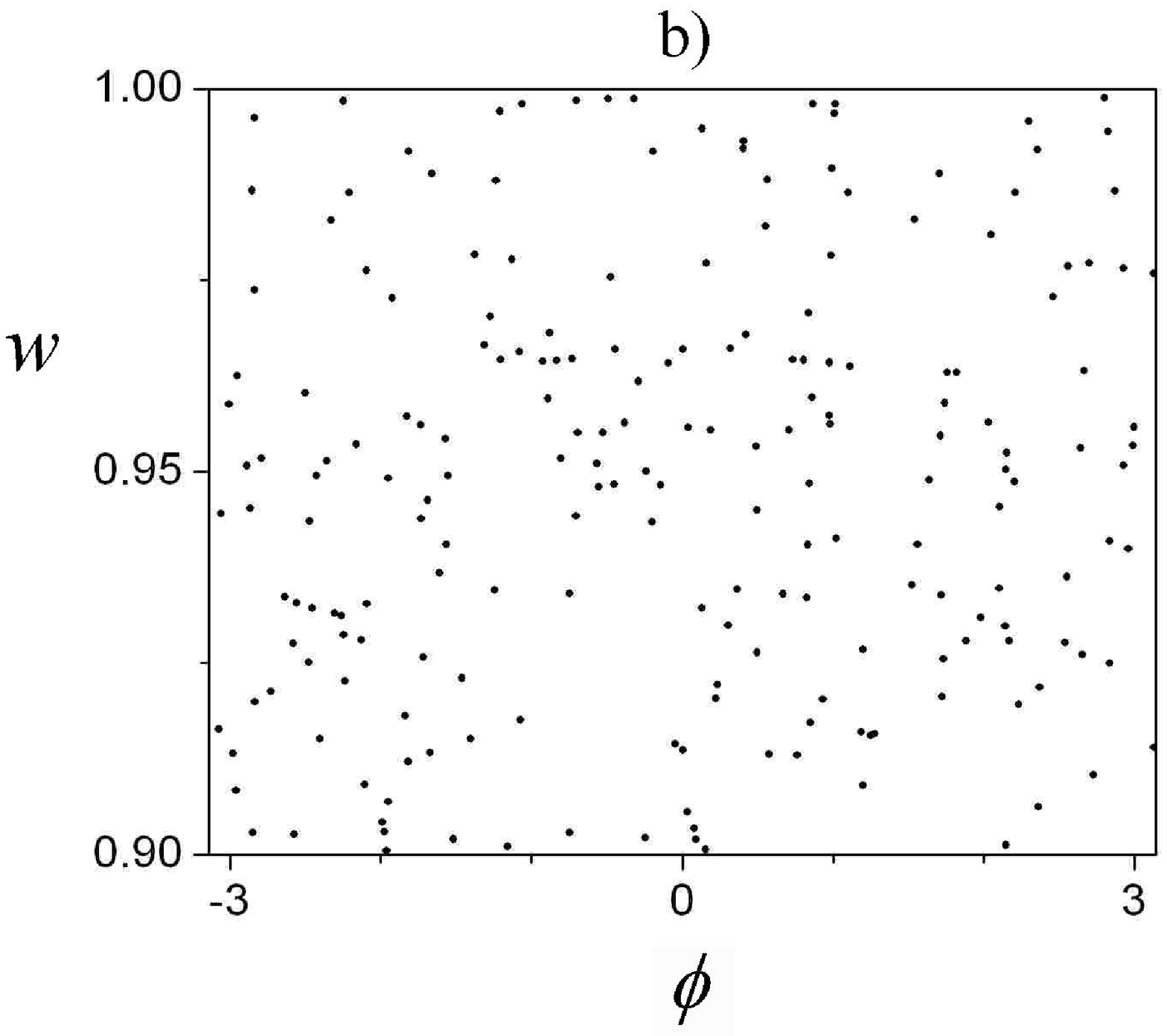}
 \caption{Stroboscopic shots of two trajectories  without/with separatrix
crossings
 (  a)/b) correspondingly), and with the same initial conditions.
Parameters:  $\eps=0.002,$ $\delta_{0}=-10$. In (a),
 value of $\delta$ was modulated as $\delta=\delta_{0}(0.75+0.25 \cos
\eps
 t)$, so no separatrix crossing occur. In (b), $\delta=\delta_{0} \cos
\eps
 t$, so multiple separatrix crossings result in diffusion of
 adiabatic invariant as described in the text.
  \label{poincare}}
\end{figure*}

An example of such diffusion is shown in Fig. \ref{poincare}b. The figure demonstrates
the result of integration along one phase trajectory over a long period of time with
$\delta$ being modulated as $\delta=\delta_{0} \cos(\tau), \quad \delta_{0}=-10$. For the
presentation, we plot the corresponding phase points on this trajectory at instances of
time when $\delta=\delta_{0}$. For the comparison, we plot the results for a trajectory
with the same initial conditions, but with  $\delta$ being modulating as
$\delta=\delta_{0}(0.75+0.25 \cos(\tau)), \quad \delta_{0}=-10$, i.e. without the
separatrix crossings (Fig. \ref{poincare}a).  The same stroboscopic shots produce smooth
curve: the adiabatic invariant is eternally conserved in this system at sufficiently
small $\eps$ \cite{AKN}.

It is interesting to note that (for system with separatrix crossings) the variations in
$\xi$ may be correlated along certain trajectories. As it was shown in \cite{NST97} in a
more general case, this results in existence of stable periodic trajectories and
stability islands in the domain of the separatrix crossings. Total measure of these
islands does not tend to zero as $\eps \rightarrow 0$, yet it is small. Along a phase
trajectory with initial conditions inside such an island, the value of the adiabatic
invariant undergoes only periodic oscillations with amplitude of order $\eps$
\cite{NST97}.

\section{Acknowledgements}

A.P.I. is supported by JSPS. The authors thank A.I. Neishtadt for
clarifying discussions. A.A.V. was supported by the RFBR grant
06-01-00117.  This work was supported also by Grants-in-Aid for
Scientific Research No. 15540381 and 16-04315 from the Ministry of
Education, Culture, Sports, Science and Technology, Japan, and
also in part by the 21st Century COE program on ``Coherent Optical
Science''.  A.P.I gratefully acknowledges short encouraging
discussions with R.Hulet and E.Arimondo, and thanks D.Feder and
A.Perali for their help during participation in BCAM05 meeting and
NQS2005 conference correspondingly.

\end{document}